\documentstyle[12pt]{article}
\newcommand{\be}{\begin{equation}}
\newcommand{\ee}{\end{equation}}
\newcommand{\bea}{\begin{eqnarray}}
\newcommand{\eea}{\end{eqnarray}}
\newcommand{\norsl}{\normalsize\sl}
\newcommand{\norsc}{\normalsize\sc}

\def\Cv{\mbox{\boldmath $C$}}

\def\qv{\mbox{\boldmath $q$}}

\begin{document}

%-------------------- Title page ----------------------------------
\begin{titlepage}

\title{ Polarized Parton Distributions in the Photon \\
and  Factorization Scheme Dependence}
\author{
\norsc  Ken SASAKI\thanks{e-mail address: sasaki@ed.ynu.ac.jp}~ and
      Tsuneo UEMATSU\thanks{e-mail address: uematsu@phys.h.kyoto-u.ac.jp} \\
\norsl  $^*$Dept. of Physics,  Faculty of Engineering, Yokohama National
University \\
\norsl  Yokohama 240-8501, JAPAN \\
\norsl  $^{\dag}$Dept. of Fundamental Sciences, FIHS, Kyoto University \\
\norsl     Kyoto 606-8501, JAPAN \\
}

\date{}
\maketitle

\begin{abstract}
{\normalsize
Spin-dependent
parton distributions in the polarized virtual photon are
investigated
 in QCD up to the next-to-leading order (NLO). In the case
$\Lambda^2 \ll P^2 \ll Q^2$, where $-Q^2$ ($-P^2$) is the mass squared of
the probe (target) photon, parton distributions can be predicted completely
up to
the NLO, but they are factorization-scheme-dependent. Parton distributions are
analyzed  in four different factorization schemes and their scheme dependence
are discussed.  Particular attentions are paid to the
axial anomaly effect on the first moments of quark parton distributions, and
also to
the large-$x$  behaviors of the parton distributions.
Gluon distribution in the virtual photon is found to be
factorization-scheme independent up to the NLO. }
\end{abstract}

\begin{picture}(5,2)(-260,-500)
\put(2.3,-55){YNU-HEPTh-99-102}
\put(2.3,-70){KUCP-144}
\put(2.3,-85){November 1999}
\put(2.3,-100){Revised December 1999}
\end{picture}

\thispagestyle{empty}
\end{titlepage}
\setcounter{page}{1}
\baselineskip 17pt
%----------------------- Text -----------------------------------

In the past few years, the accuracy of the experimental data on the
spin dependent structure function $g_1$ of the nucleon has been
significantly improved~\cite{Nucleon}.
Using these experimental data together with the already existing world data,
it is now possible to extract polarized parton (i.e., quark and gluon)
distributions in the nucleon in the framework of perturbative QCD.
In fact several
groups~\cite{AnalysisGRSVGS,AnalysisABFR,AnalysisLSS,AnalysisEX} have
carried out the next-to-leading order
(NLO) QCD analysis on the  polarized parton distributions in the nucleon
recently.
These parton distributions may be used for predicting the behaviors of other
processes
such as polarized Drell-Yan reactions and polarized semi-inclusive deep
inelastic
scatterings, and etc.
The first moments of  polarized parton distributions are particularly
interesting due to their  relevance for the spin structure of the
nucleon~\cite{GluonCon,BQ}, where the axial anomaly plays an important
role~\cite{JK}.
However, at the NLO and beyond in perturbative QCD, parton distribution
functions become
dependent on the factorization (or renormalization) scheme employed.
%Since we need to resort to some assumptions in order to extract parton
%distributions in the nucleon from experimental data, it is rather difficult
%to see the features of each factorization scheme adopted.

Recently, the first moment of the real photon structure function $g_1^\gamma$
has attracted attention in the literature \cite{ET,FS,BBS}. More recently,
the present authors investigated the spin-dependent
structure function $g_1^{\gamma}(x, Q^2, P^2)$ of the  virtual photon\footnote{
The NLO QCD analysis on $g_1^{\gamma}$ for the real photon target was made by
Stratmann and Vogelsang~\cite{SV}. The  leading order QCD correction to
$g_1^{\gamma}$
was first studied by one of the authors~\cite{KS1}. } in the
NLO in QCD~\cite{SU}. The advantage in  studying the virtual photon target
is that, in the case
$\Lambda^2 \ll P^2 \ll Q^2$, where $-Q^2$ ($-P^2$) is the mass squared of
the probe (target) photon, and $\Lambda$ is the QCD scale parameter, we can
calculate the whole structure function up to the NLO by the perturbative
method~\cite{UW1,UW2}, in
contrast to the case  of the real photon target where there exist
non-perturbative pieces in the NLO.

In this paper we analyze the polarized parton distributions in the
virtual photon target.  The  behaviors of the parton distributions can be
predicted entirely up to the NLO,
but, of course, they are factorization-scheme-dependent.
We carry out our analysis in four different factorization schemes,
(i) $\overline{\rm MS}$, (ii) CI (chirally invariant) (it is also called as
JET)~\cite{HYC,MT},   (iii) AB (Adler-Bardeen)~\cite{BFR}, and (iv) OS
(off-shell)~\cite{BFR},
and see  how the parton distributions change in each scheme. In particular,
we examine in
detail the axial anomaly effect on the first moments and the large-$x$
behaviors of the
parton distributions in each scheme. Gluon distribution in the virtual
photon is found to be
factorization-scheme independent up to the NLO.

Let $\Delta q^{\gamma}_S(x,Q^2,P^2)$, $\Delta q^{\gamma}_{NS}(x,Q^2,P^2)$,
$\Delta G^{\gamma}(x,Q^2,P^2)$,
$\Delta \Gamma^{\gamma}(x,Q^2,P^2)$ be the flavor singlet-,
non-singlet-quark,  gluon, and  photon distribution functions, respectively,
in the longitudinally polarized virtual photon with mass $-P^2$.
In the leading order of the electromagnetic coupling constant,
$\alpha=e^2/4\pi$,
$\Delta \Gamma^{\gamma}$ does not evolve with $Q^2$ and is set to be
$\Delta \Gamma^{\gamma}(x,Q^2,P^2)=\delta(1-x)$.
In terms of these parton distribution
functions, the polarized virtual photon  structure function
$g_1^{\gamma}(x, Q^2, P^2)$ is expressed in the QCD improved parton model
as~\cite{SU}
\bea
&& \hspace{-0.8cm} g_1^{\gamma}(x, Q^2,P^2)
   = \int^1_x \frac{dy}{y}~\biggl\{
   \Delta q^{\gamma}_S (y,Q^2,P^2)~\Delta C^{\gamma}_S (\frac{x}{y}, Q^2)
                 + \Delta G^{\gamma} (y,Q^2,P^2)~\Delta C^{\gamma}_G
(\frac{x}{y},
Q^2)~   \nonumber  \\
  & &  \hspace{4.5cm} + \Delta q^{\gamma}_{NS}
(y,Q^2,P^2) ~\Delta C^{\gamma}_{NS} (\frac{x}{y}, Q^2)
\biggr\}   + \Delta C^{\gamma}_{\gamma} (x, Q^2) \label{Solg}
\eea
where $\Delta C^{\gamma}_S$($\Delta C^{\gamma}_{NS}$), $\Delta
C^{\gamma}_G$, and
$\Delta C^{\gamma}_{\gamma}$ are the coefficient functions  corresponding to
singlet(non-singlet)-quark, gluon, and photon, respectively,  and are
independent of
$P^2$. The Mellin moments of $g_1^{\gamma}$ is written  as
\be
 g_1^{\gamma}(n,Q^2,P^2)=\Delta\Cv^{\gamma}(n,Q^2) \cdot
\Delta\qv^{\gamma}(n,Q^2,P^2)
\label{gonegamma} \\
\ee
where
\bea
   \Delta\Cv^{\gamma}(n,Q^2)&=&(\Delta C^{\gamma}_S~,~
 \Delta C^{\gamma}_G~,~\Delta C^{\gamma}_{NS}~,~
  \Delta C^{\gamma}_{\gamma}) \nonumber  \\
\Delta\qv^{\gamma}(n,Q^2,P^2)&=&(\Delta q_S^{\gamma}~,~
\Delta G^{\gamma}~,~ \Delta q_{NS}^{\gamma}~, ~\Delta \Gamma^{\gamma})
\nonumber
\eea
and the matrix notation is implicit.

The parton distribution $\Delta\qv^{\gamma}$ satisfies the inhomogeneous
evolution equation \cite{FP1,GR,GRV,FP2}.
The explicit expressions of $\Delta q^\gamma_S$,
$\Delta G^\gamma$, and $\Delta q^\gamma_{NS}$ up to the NLO are easily derived
from Eq.(4.46) of Ref.\cite{SU}.
They are given\footnote{
We use the same notations as in Ref.\cite{SU}, except that
the symbol $\Delta$ has been appended to all the  spin-dependent
anomalous dimensions and coefficient functions. } in
terms of  one-(two-) loop hadronic anomalous dimensions
$\Delta\gamma_{ij}^{(0),n}$
($\Delta\gamma_{ij}^{(1),n}$) ($i,j=\psi, G$) and
$\Delta\gamma_{NS}^{(0),n}$
($\Delta\gamma_{NS}^{(1),n}$), one-(two-) loop anomalous dimensions\nolinebreak
$\Delta K_i^{(0),n}$ ($\Delta K_i^{(1),n}$) ($i=\psi, G, NS$) which
represent the
mixing between photon and three hadronic operators $R_i^n$ ($i=\psi, G,
NS$), and
finally
$A_i^n$, the one-loop photon matrix elements  of  hadronic operators
renormalized at
$\mu^2= P^2(=-p^2)$,
\be
  \langle \gamma (p) \mid R_i^n (\mu) \mid \gamma (p) \rangle \vert_{\mu^2=
P^2}
=\frac{\alpha}{4\pi}  A_i^n     \qquad  (i=\psi, G, NS)~.   \label{Initial}
\ee

Although $g_1^{\gamma}$ is a
physical quantity and thus unique, there remains a freedom in the
factorization of
$g_1^{\gamma}$  into $\Delta\Cv^{\gamma}$ and $\Delta\qv^{\gamma}$.
Given the formula Eq.(\ref{gonegamma}),
we can always redefine $\Delta\Cv^{\gamma}$ and $\Delta\qv^{\gamma}$ as follows
\cite{FP1}:
\bea
 \Delta\Cv^{\gamma}(n,Q^2)\rightarrow \Delta{\Cv}^{\gamma}(n,Q^2)\vert_a
      &\equiv& \Delta\Cv^{\gamma}(n,Q^2)Z^{-1}_a(n,Q^2)   \nonumber  \\
 \Delta\qv^{\gamma}(n,Q^2,P^2)  \rightarrow \Delta{\qv}(n,Q^2,P^2)\vert_a
 &\equiv& Z_a(n,Q^2)~ \Delta\qv^{\gamma}(n,Q^2,P^2)
\eea
where $\Delta{\Cv}^{\gamma}\vert_a  $ and $ \Delta
{\qv}\vert_a $  correspond to the quantities in a new factorization
scheme-$a$.  Note that the coefficient functions and anomalous
dimensions are closely connected under factorization. We will study
the factorization scheme dependence of parton distribution up to the NLO, by
which we mean that a scheme transformation  for the coefficient functions is
considered up to the one-loop order, since a NLO prediction for
$g_1^{\gamma}$ is given by the one-loop coefficient functions and
anomalous dimensions up to the two-loop order.

The most general form of a transformation for the coefficient functions
in one-loop order, from $\overline {\rm MS}$ scheme to
a new factorization scheme-$a$, is given by
\bea
    \Delta C_{S,~a}^{\gamma,~ n}&=&
\Delta C_{S,~\overline{\rm MS}}^{\gamma,~ n}~
-<e^2>\frac{\alpha_s}{2\pi}~w_S(n,a) \nonumber \\
\Delta C_{G,~a}^{\gamma,~ n}&=&\Delta C_{G,~\overline {\rm MS}}^{\gamma,~ n}~
-<e^2>\frac{\alpha_S}{2\pi}~z(n,a)  \nonumber\\
\Delta C_{NS,~a}^{\gamma,~ n}&=&\Delta C_{NS,~\overline {\rm MS}}^{\gamma,~ n}~
-\frac{\alpha_s}{2\pi}~w_{NS}(n,a) \label{Coeffgamma} \\
 \Delta C_{\gamma,~a}^{\gamma,~ n}&=&
\Delta C_{\gamma,~\overline{\rm MS}}^{\gamma,~ n}~
-\frac{\alpha}{\pi}~3<e^4>{\hat z}(n,a) \nonumber
\eea
where  $<e^2>=\sum_i e^2_i/N_f$, $<e^4>=\sum_i e^4_i/N_f$, with
$N_f$ being the number of flavors of active quarks and $e_i$ being the
electric charge of $i$-flavor-quark.
The $z(n,a)$ (${\hat z}(n,a)$) term tells how much of the
QCD (QED) axial-anomaly effect is transferred to the coefficient function
in the new factorization scheme.  Note that
$\Delta C_{\gamma,~\overline {\rm MS}}^{\gamma,~ n}$ has been obtained  from
$\Delta C_{G,~\overline {\rm MS}}^{\gamma,~ n}$,  with changes:
%$\frac{\alpha_S}{2\pi}\rightarrow
%\frac{2\alpha}{\alpha_S}\times\frac{\alpha_S}{2\pi}$~,
$\alpha_S/{2\pi}\rightarrow
(2\alpha/{\alpha_S})\times (\alpha_S/{2\pi})$~,
$<e^2>\rightarrow 3<e^4>$,  and $3$ is the number of colors.
Once the relations (\ref{Coeffgamma}) between the coefficient functions in the
$a$-scheme and $\overline {\rm MS}$ scheme are given, we can derive
corresponding
transformation rules from $\overline {\rm MS}$ scheme to $a$-scheme for the
relevant
two-loop anomalous dimensins. We find\footnote{For detailed derivation
of the transformation rules, \
see Ref.\cite{SU2}.}
\bea
  \Delta \gamma^{(1),n}_{\psi\psi,~a}&=&\Delta
\gamma^{(1),n}_{\psi\psi,~\overline{\rm MS}}
   + 2 z(n,a)~\Delta \gamma^{(0),n}_{G\psi} +4 \beta_0 w_S(n,a)\nonumber\\
 \Delta \gamma^{(1),n}_{\psi G,~a}&=&\Delta \gamma^{(1),n}_{\psi
G,~\overline {\rm MS}}
   + 2 z(n,a)~\Bigl[ \Delta \gamma^{(0),n}_{GG} -
  \Delta \gamma^{(0),n}_{\psi\psi} + 2\beta_0 \Bigr]  \nonumber  \\
& &+ 2w_S(n,a) \Delta \gamma^{(0),n}_{\psi G}\nonumber\\
 \Delta \gamma^{(1),n}_{G\psi,~a}&=&\Delta \gamma^{(1),n}_{G\psi,
~\overline{\rm MS}}
- 2w_S(n,a) \Delta \gamma^{(0),n}_{G\psi}\nonumber\\
 \Delta \gamma^{(1),n}_{GG,~a}&=&\Delta \gamma^{(1),n}_{GG,~\overline {\rm MS}}
 - 2 z(n,a) \Delta \gamma^{(0),n}_{G\psi} \nonumber\\
    \Delta \gamma^{(1),n}_{NS,~a}&=&\Delta
\gamma^{(1),n}_{NS,~\overline {\rm MS}}
+4 \beta_0 w_{NS}(n,a)\label{Transformation}\\
  \Delta K^{(1),n}_{S,~a}&=& \Delta K^{(1),n}_{S,~\overline {\rm MS}}
+  2w_S(n,a) \Delta K^{(0),n}_S
+ 4 {\hat z}(n,a)3<e^2> \Delta \gamma^{(0),n}_{\psi\psi}  \nonumber\\
 \Delta K^{(1),n}_{G,~a}&=& \Delta K^{(1),n}_{G,~\overline {\rm MS}}
+ 4 {\hat z}(n,a)3<e^2> \Delta \gamma^{(0),n}_{G\psi}\nonumber \\
 \Delta K^{(1),n}_{NS,~a}&=& \Delta K^{(1),n}_{NS,~\overline {\rm MS}}
 +  2w_{NS}(n,a) \Delta K^{(0),n}_{NS}  \nonumber \nonumber \\
& & + 4 {\hat z}(n,a)3(<e^4>-<e^2>^2) \Delta \gamma^{(0),n}_{NS} \nonumber
\eea
where $\beta_0=11-\frac{2}{3}N_f$ is the one-loop coefficient of the
QCD beta function.

Since the one-loop photon matrix elements of the hadronic operators,
$A^n_{\psi}$ and  $A^n_{NS}$ in Eq.(\ref{Initial}),  are related to each
other as $A^n_{NS}= A^n_{\psi}(<e^4>-<e^2>^2)/<e^2>$
and the sum $(\Delta C_\gamma^{\gamma,~n}/\frac{\alpha}{4\pi}+<e^2>
A^n_{\psi}+A^n_{NS})$
is factorization-scheme-independent in one-loop order\cite{SU},
we find from Eq.(\ref{Coeffgamma})
 \bea
    A_{\psi,~a}^n&=&A_{\psi,~\overline {\rm MS}}^n +12<e^2>{\hat z}(n,a)
\nonumber\\
 A_{G,~a}^n&=&A_{G,~\overline {\rm MS}}^n =0   \label{TransForMat}\\
 A_{NS,~a}^n&=&A_{NS,~\overline {\rm MS}}^n
+12(<e^4>-<e^2>^2)~{\hat z}(n,a)~.  \nonumber
\eea
Note that $A^n_G=0$ in one-loop order.

It is possible to choose $z(n,a)$ and ${\hat z}(n,a)$ arbitrarily.  But, here,
we are interested in the QCD and QED anomaly effects on the parton
distributions in the virtual photon, and both QCD and QED anomalies originate
from the similar triangle diagrams. Therefore, we take in the following
$z(n,a)={\hat z}(n,a)$.   With this choice, the relation
between the one-loop gluon and photon coefficient functions
\be
    \Delta B_{\gamma}^n=\frac{2}{N_f}\Delta B_G^n~,
\ee
holds not only in the $\overline {\rm MS}$ scheme but also in the $a$-scheme,
where $\Delta B_{\gamma}^n$ and $\Delta B_G^n$ are defined as
\bea
 \Delta C_G^{\gamma,~n}&=&<e^2>\Bigl(\frac{\alpha_s}{4\pi}\Delta B_G^n
+{\cal O}
(\alpha_s^2)~\Bigr) \nonumber\\
 \Delta C^{\gamma,~n}_\gamma&=&\frac{\alpha}{4\pi}3N_f<e^4>
     \Bigl(\Delta B_{\gamma}^n +{\cal O}(\alpha_s)~\Bigr)~.
\eea
Also in one-loop order  we have $w_S(n,a)=w_{NS}(n,a)$. Thus from now on,
we set
${\hat z}(n,a)=z(n,a)$ and $w_S(n,a)=w_{NS}(n,a)\equiv w(n,a)$.

%\bigskip

Now let us discuss the features of several factorization schemes.

(i) [The $\overline {\rm MS}$ scheme]
\ \ This is the only scheme in which both relevant one-loop coefficient
functions and
two-loop anomalous dimensions were actually calculated~\cite{KMMSU,BQ,MvN,V}.
In fact there still remain ambuguities in the $\overline {\rm MS}$ scheme,
depending on
how to handle $\gamma_5$ in $n$ dimensions.
The $\overline {\rm MS}$ scheme we call here is the one due to Mertig and van
Neerven~\cite{MvN} and  Vogelsang~\cite{V}, in which the first moment of the
non-singlet
quark operator  vanishes, corresponding to the conservation of the
non-singlet axial
current. Indeed we have $\Delta \gamma^{(1),n=1}_{NS,~\overline {\rm MS}}=0$.
Explicit expressions of the relevant one-loop coefficient functions and
two-loop anomalous dimensions can be found, for example, in Appendix
of  Ref.~\cite{SU}. In the  $\overline {\rm MS}$ scheme, the QCD (QED)
axial anomaly
resides  in the quark distributions and not in the gluon (photon)
coefficient function
\cite{BQ,HYC}.
In fact we observe
\be
\Delta \gamma^{(1),n=1}_{\psi\psi,~\overline {\rm MS}}=
24C_FT_f \neq 0~, \qquad
\Delta B_{G,~\overline {\rm MS}}^{n=1}=\Delta B_{\gamma,~
\overline{\rm MS}}^{n=1}=0~.
\ee
Also the first moment of the one-loop photon matrix element of quark operators
gains the non-zero values, i.e.,
\be
A_{\psi,~\overline {\rm MS}}^{n=1}=\frac{<e^2>}{<e^4>-<e^2>^2}~
A_{NS,~\overline {\rm MS}}^{n=1}=-12<e^2>N_f  \label{PhotonMSn=1}
\ee
which is due to the QED axial anomaly.

(ii) [The chirally invariant (CI)  scheme]
\ \ In this scheme the factorization of the photon-gluon (photon-photon) cross
section into the hard and soft parts is made so that  chiral symmetry is
respected and
all the anomaly effects are absorbed into the gluon (photon) coefficient
function\cite{HYC,MT}. Thus the spin-dependent quark distributions in
the CI
scheme are  anomaly-free. In particular, we have
\bea
 \Delta B_{G,~{\rm CI}}^{n=1}&=&-2N_f~, \qquad  \Delta
\gamma^{(1),n=1}_{\psi\psi,~{\rm CI}}=0
\label{CIlike}\\
\Delta B_{\gamma,~{\rm CI}}^{n=1}&=&-4~, \qquad
\quad A_{\psi,~{\rm CI}}^{n=1}= A_{NS,~{\rm CI}}^{n=1}=0~.
\eea
The transformation from
$\overline {\rm MS}$ scheme to CI scheme is achieved by
\be
   w(n,a={\rm CI})=0~, \qquad
   z(n,a={\rm CI})=2N_f\frac{1}{n(n+1)}~.  \label{TransCI}
\ee
It has been argued by Cheng \cite{HYC} and
M\"{u}ller and Teryaev \cite{MT} that
the $x$-dependence of the axial-anomaly effect is uniquely fixed and that its
$x$-behavior  leads to the transformation rule (\ref{TransCI}) and thus to
the CI
scheme.

(iii) [The Adler-Bardeen (AB)  scheme]
\ \ Ball, Forte and Ridolfi~\cite{BFR} proposed
several CI-like schemes in which features of the CI scheme
(CI-relations in Eq.(\ref{CIlike}))
are kept intact. One of them is the Adler-Bardeen (AB) scheme which was
introduced by
requiring that the change from the $\overline {\rm MS}$ scheme to this scheme
be
independent of $x$, so that the large and small $x$ behavior of the gluon
(photon) coefficient function is unchanged. In moment space we have
\be
   w(n,a={\rm AB})=0~,\qquad
   z(n,a={\rm AB})=N_f\frac{1}{n}~.
\ee

(iv) [The off-shell (OS)  scheme]
\ \ In this scheme~\cite{BFR} we renormalize operators while keeping the
incoming
particle
off-shell, $p^2\neq 0$, so that at renormalization (factorization) point
$\mu^2=-p^2$, the finite terms vanish.  This is exactly the same as ``the
momentum
subtraction scheme" which was used some time ago to calculate, for instance,
the polarized quark and gluon coefficient
functions~\cite{KMSU,JK}\footnote{In fact, the author of
Ref.~\cite{JK} treated the $n=1$ moment of the gluon
coefficient function differently from other moments \cite{SU}.}.
The CI-relations in Eq.(\ref{CIlike}) hold in the OS scheme, since
the axial anomaly appears as a finite term in the calculation of the
triangle graph for $j_5^{\mu}$ between external gluons (photons) and
the finite term is thrown away in this scheme.  The transformation  from the
$\overline{\rm MS}$ scheme  to the OS scheme is made by choosing
\bea
   w(n,a={\rm OS})&=& C_F \biggl\{\Bigl[S_1(n)\Bigr]^2+3S_2(n)-
S_1(n)\Bigl(\frac{1}{n}- \frac{1}{(n+1)}  \Bigr)  \nonumber  \\
& & \qquad \qquad-\frac{7}{2} +\frac{2}{n}
-\frac{3}{n+1}-\frac{1}{n^2}+\frac{2}{(n+1)^2}\biggr\}\nonumber \\
   z(n,a={\rm OS})&=&N_f\biggl\{ -\frac{n-1}{n(n+1)}S_1(n)+ \frac{1}{n}
+\frac{1}{n^2}-\frac{4}{(n+1)^2} \biggr\}~.
\eea
It is noted that in the OS scheme we have
$ A^n_{\psi,~{\rm OS}}=A^n_{NS,~{\rm OS}}=0 $
for all $n$.

Now we examine the factoraization scheme dependence of the
polarized parton distributions in virtual photon.
The two-loop anomalous dimensions of the spin-dependent operators and
one-loop photon matrix elements of the hadronic operators in the
$\overline{\rm MS}$
scheme are already known.
Corresponding quantities in a particular scheme are obtained through the
transformation
rules in Eq.(\ref{Transformation}). Using these quantities, we get the NLO
predictions
for the moments of polarized  parton distributions in
virtual photon in a particular factorization scheme.  We find that the
gluon
distribution is factorization-scheme  independent up to the NLO
\footnote{
By NLO we mean that we consider a general scheme transformation
for the coefficient functions up to the one-loop order, which is given by
Eq.(\ref{Coeffgamma}).}
,
\be
   \Delta G^\gamma(n,Q^2,P^2)\vert_a =\Delta
G^\gamma(n,Q^2,P^2)\vert_{\overline {\rm MS}}~,
\ee
where $a$ means CI, AB, OS,  or any other factorization scheme.
This can be seen from the direct calculation or from the notion that, up to
the NLO,
$\Delta G^\gamma\vert_a$ satisfies the same evolution equation as
$\Delta G^\gamma\vert_{\overline {\rm MS}}$  with the same initial condition
at $Q^2=P^2$, namely,~ $\Delta G^\gamma(n,P^2,P^2)\vert_a =\Delta
G^\gamma(n,P^2,P^2)\vert_{\overline {\rm MS}}=0$.

%\bigskip
%\noindent
(1) [The first moments]\ \  For all three factorization schemes,
$a={\rm CI,AB,OS}$,
we have
\be
   w(n=1,a)=0,~\quad z(n=1,a)=N_f   \qquad {\rm for}\ a={\rm CI,AB,OS}
\ee
These schemes, therefore, give the same first moments for the parton
distributions.
In fact, from Eqs.(\ref{TransForMat}) and (\ref{PhotonMSn=1}) we find~
$ A_{\psi,~a}^{n=1}=A_{NS,~a}^{n=1}=0$~.
This leads to
\be
\Delta q_S^\gamma(n=1,Q^2,P^2)\vert_a =\Delta
q_{NS}^\gamma(n=1,Q^2,P^2)\vert_a
=0   \label{QSCI}
\ee
in the NLO for $a={\rm CI,AB,OS}$. In these schemes, the axial anomaly effects
 are
transfered to the gluon and photon coefficient functions. On the other hand,
in
$\overline {\rm MS}$ scheme we  obtain\footnote{The detailed derivation will be
reported
elsewhere\cite{SU2}}
\bea
  \Delta q_S^{\gamma}(n=1,Q^2,P^2)\vert_{\overline {\rm MS}}&=&
\Bigl[-\frac{\alpha}{\pi}~3 <e^2>N_f  \Bigr]
\left\{1-\frac{2}{\beta_0}\frac{\alpha_s(P^2)-\alpha_s(Q^2)}{\pi}
 N_f\right\}  \nonumber \label{QSMS}\\
\Delta q_{NS}^{\gamma}(n=1,Q^2,P^2)\vert_{\overline {\rm MS}}&=&
\Bigl[-\frac{\alpha}{\pi}~3 \Bigl( <e^4>-<e^2>^2 \Bigr) N_f  \Bigr]
\left\{1+{\cal O}(\alpha_s^2) \right\}  \label{QNSMS}
\eea
For gluon distribution, we have
\bea
\Delta G^\gamma(n=1,Q^2,P^2)\vert_{\overline {\rm MS}}&=&
\Delta G^\gamma(n=1,Q^2,P^2)\vert_{a}  \nonumber  \\
&=&\frac{12\alpha}{\pi\beta_0}
<e^2>N_f~\frac{\alpha_s(Q^2)-\alpha_s(P^2)}{\alpha_s(Q^2)}~.
\eea

The polarized structure function $ g_1^{\gamma}(x,Q^2,P^2)$ of the virtual
photon
satisfies the following sum rule\cite{SU,NSV}:
\bea
\int_0^1dx g_1^\gamma(x,Q^2,P^2)
&=&-\frac{3\alpha}{\pi}<e^4>N_f
\left(1-\frac{\alpha_s(Q^2)}{\pi}\right)       \nonumber\\
&+&\frac{6\alpha}{\pi\beta_0}\Bigl[<e^2>N_f \Bigr]^2~
\frac{\alpha_s(P^2)-\alpha_s(Q^2)}{\pi}
+{\cal O}(\alpha_s^2). \label{Oalpha}
\eea
This sum rule is of course the factorization-scheme independent. Now we
examine
how the scheme-dependent parton distributions contribute to this sum rule.
In the
CI-like schemes ($a={\rm CI,AB,OS}$), the first moment of the quark
distributions
vanish in the NLO, and thus the contribution to the sum rule come from the
gluon and
photon  distributions. Since
\bea
\Delta C_{G,~a}^{\gamma,~n=1}&=&-<e^2>\frac{\alpha_s(Q^2)}{2\pi}N_f
\nonumber\\
\Delta C_{\gamma,~a}^{\gamma,~n=1}&=&-\frac{3\alpha}{\pi}<e^4>N_f
\left(1-\frac{\alpha_s(Q^2)}{\pi}\right)
\eea
we see that $[\Delta C_{G,~a}^{\gamma,~n=1}\Delta
G^\gamma(n=1,Q^2,P^2)\vert_{a}+
\Delta C_{\gamma,~a}^{\gamma,~n=1}]$ leads to the result (\ref{Oalpha}).
On the other hand, in the $\overline {\rm MS}$ scheme, the one-loop gluon and
photon
coefficient functions vanish, $\Delta B_{G,~\overline {\rm MS}}^{n=1}=\Delta
B_{\gamma,~\overline {\rm MS}}^{n=1}=0$ and, therefore, the sum rule is derived
from the
quark contributions. Indeed  we have in one-loop order
\be
\frac{1}{<e^2>}\Delta C_{S,~\overline {\rm MS}}^{\gamma,~ n=1}=
\Delta C_{NS,~\overline {\rm MS}}^{\gamma,~
n=1}=\left(1-\frac{\alpha_s(Q^2)}{\pi}\right)
\ee
and find that
\be
\Delta C_{S,~\overline {\rm MS}}^{\gamma,~ n=1}~
 \Delta q_S^{\gamma}(n=1,Q^2,P^2)\vert_{\overline {\rm MS}}~+
\Delta C_{NS,~\overline {\rm MS}}^{\gamma,~ n=1}~
\Delta q_{NS}^{\gamma}(n=1,Q^2,P^2)\vert_{\overline {\rm MS}}
\ee
leads to the same result.

It is interesting to note that the sum rule (\ref{Oalpha}) is the
consequence of the
axial anomaly and that in the CI-like schemes the anomaly effect resides in
the gluon contribution while, in the $\overline {\rm MS}$, in the quark
contributions.
Furthermore, the first term of the sum rule (\ref{Oalpha}) is coming from
the QED
axial anomaly and the second is from the QCD axial anomaly.

%\bigskip
%\noindent
(2) [behaviors near~ $x=1$]\ \ The behaviors of parton distributions near~
$x=1$
are governed by the large-$n$ limit of those moments. In the leading order
(LO),
parton distributions are factorization-scheme independent. For large $n$,~
$\Delta q_S^{\gamma}(n,Q^2,P^2)\vert_{\rm LO}$ and $\Delta
q_{NS}^{\gamma}(n,Q^2,P^2)\vert_{\rm LO}$ behave as $1/(n~{\rm ln}~n)$,while
$\Delta G^{\gamma}(n,Q^2,P^2)\vert_{\rm LO}\propto 1/(n~{\rm ln}~n)^2$.
Thus in $x$ space,  the parton distributions vanish for $x \rightarrow 1$.
In fact we find
\bea
 \Delta q_S^{\gamma}(x,Q^2,P^2)\vert_{\rm LO}&\approx&
\frac{\alpha}{4\pi}\frac{4\pi}{\alpha_s(Q^2)}
N_f<e^2>\frac{9}{4}~\frac{-1}{{\rm ln}~(1-x)} \\
\Delta G^{\gamma}(x,Q^2,P^2)\vert_{\rm LO}&\approx&
\frac{\alpha}{4\pi}\frac{4\pi}{\alpha_s(Q^2)}
N_f<e^2>\frac{1}{2}~\frac{-{\rm ln}~x}{{\rm ln}^2~(1-x)}~. \nonumber
\eea
The behaviors of $\Delta q_{NS}^{\gamma}(x,Q^2,P^2)$ for $x \rightarrow 1$,
both in the LO and NLO, are always
given by the corresponding expressions for $\Delta q_{S}^{\gamma}(x,Q^2,P^2)$
with the replacement of the charge factor $<e^2>$ with $(<e^4>-<e^2>^2)$.

In the $\overline {\rm MS}$ scheme,
the moment $ \Delta q_S^{\gamma}(n,Q^2,P^2)\vert_{{\rm NLO},~\overline {\rm
MS}}$
behaves as $({\rm ln}~n)/n$ in large $n$ limit, while
$\Delta G^{\gamma}(n,Q^2,P^2)\vert_{{\rm NLO},~\overline {\rm MS}}\propto
1/n^2$.  Thus we have near $x=1$
\bea
   \Delta q_S^{\gamma}(x,Q^2,P^2)\vert_{{\rm NLO},~\overline {\rm MS}}&\approx&
\frac{\alpha}{4\pi} N_f<e^2>6~\Bigl[-{\rm ln}(1-x)\Bigr]\label{NLOMSbar}\\
\Delta G^{\gamma}(x,Q^2,P^2)\vert_{{\rm NLO},~\overline {\rm MS}}&\approx&
\frac{\alpha}{4\pi} N_f<e^2>3~\Bigl[-{\rm ln}~x\Bigr] ~. \nonumber
\eea
It is remarkable that, in the $\overline {\rm MS}$ scheme, quark parton
distributions,
$\Delta q_S^{\gamma}(x)\vert_{{\rm NLO},~\overline {\rm MS}}$ and
$\Delta q_{NS}^{\gamma}(x)\vert_{{\rm NLO},~\overline {\rm MS}}$
diverge as $[-{\rm
ln}(1-x)]$ for $x\rightarrow 1$. Recall that
$\Delta G^{\gamma}\vert_{\rm NLO}$ is
scheme-independent.
The NLO quark distributions in the CI and AB schemes also diverge
as $x\rightarrow 1$. In fact we obtain for large $x$
\bea
\Delta q_S^{\gamma}(x,Q^2,P^2)\vert_{{\rm NLO},~{\rm CI}}&\approx&
\frac{\alpha}{4\pi} N_f<e^2>6~\Bigl[-{\rm ln}(1-x)\Bigr]\label{NLOCI}\\
\Delta q_S^{\gamma}(x,Q^2,P^2)\vert_{{\rm NLO},~{\rm AB}}&\approx&
\frac{\alpha}{4\pi} N_f<e^2>6~\Bigl[-{\rm ln}(1-x) +2\Bigr]~. \label{NLOAB}
\eea

On the other hand, the OS scheme gives quite different behaviors near $x=1$ for
the quark distributions. We find that the moment $\Delta
q_{S}^\gamma(n,Q^2,P^2)\vert_{\rm NLO}$ in the OS scheme behaves as $1/n$
for large $n$. Thus, in $x$ space,
$\Delta q_{S}^\gamma(x,Q^2,P^2)\vert_{{\rm NLO},~{\rm OS}}$
does not diverge for $x \rightarrow 1$ and approaches a constant value:
\be
\Delta q_{S}^\gamma(x,Q^2,P^2)\vert_{{\rm NLO},~{\rm OS}}
 \longrightarrow \frac{\alpha}{4\pi}N_f <e^2>
\Bigl[\frac{69}{8} + \frac{3}{4}N_f \Bigr]~. \label{NLOOS}
\ee

We can show\cite{SU2} that, as $x \rightarrow 1$, the polarized virtual
photon structure
function $g_1^{\gamma}(x, Q^2, P^2)$ in the NLO approaches a constant value
\be
  \kappa=\frac{\alpha}{4\pi}N_f<e^4>\Bigl[-\frac{51}{8}+\frac{3}{4}N_f \Bigr]
\label{kappa}
\ee
and the result is factorization-scheme independent. In the
$\overline{\rm MS}$, CI, AB schemes, the NLO quark parton distributions
and hence their contributions to $g_1^{\gamma}(x,Q^2,P^2)$ diverge as
$[-{\rm ln}(1-x)]$ for $x\rightarrow 1$. However, the one-loop photon
coefficient
function $\Delta C^{\gamma}_{\gamma}(x)$ in these schemes also diverges as
$[-{\rm ln}(1-x)]$ with the opposite sign and the sum becomes finite.
On the other hand, in the OS scheme, both the quark distributions and photon
coefficient function remain  finite as $x \rightarrow 1$. Therefore, as far
as the
large $x$-behaviors of quark parton distributions and photon (also gluon)
coefficient functions are concerned, the OS scheme is more appropriate than
other
schemes in the sence that they remain finite.

The constant value $\kappa$  in Eq.(\ref{kappa}) is
negative\footnote{The constant value $\kappa$ coincides exactly with the one
given in
Eq.(4.39) of Ref.\cite{KS2}, which was derived as the large $n$ limit of the
moment of
the NLO term $b_2(x)$ for the unpolarized structure
function $F_2^{\gamma}$~\cite{BB}.} unless
$N_f\geq 9$. Consequently, it seems superficially that QCD with 8 flavors or
less
predicts that the structure  function $g_1^{\gamma}(x,Q^2,P^2)$ turns out to be
negative for $x$ very close to 1,  since the leading term
$g_1^{\gamma}(x,Q^2,P^2)\vert_{\rm LO}$ vanishes as $x \rightarrow 1$.
But the fact is that $x$ cannot reach exactly one. The constraint
$(p+q)^2\geq 0$ gives $x\leq x_{max}=\frac{Q^2}{Q^2+P^2}$, and we find
\be
g_1^{\gamma}(x=x_{max},Q^2,P^2)\vert_{\rm LO}
\quad >\quad \frac{\alpha}{4\pi}N_f<e^4>\beta_0
\ee
and the sum $g_1^{\gamma}(x=x_{max},Q^2,P^2)\vert_{\rm LO+NLO}$
is indeed positive.

%\bigskip
%\noindent
(3) [Numerical analysis]\ \
The parton distribution functions are recovered from the moments by the
inverse Mellin
transformation. In Fig. 1 we plot the factorization scheme dependence of the
singlet quark
distribution  $\Delta q_S^{\gamma}(x,Q^2,P^2)$ beyond the LO in units of $(3
N_f <e^2>\alpha
/\pi){\rm ln}(Q^2/P^2)$. We have taken $N_f=3$, $Q^2=30~ {\rm
GeV}^2$, $P^2=1~ {\rm GeV}^2$,  and the QCD scale parameter $\Lambda=0.2~
{\rm GeV}$. All three
CI-like (i.e., CI, AB and OS) lines cross the $x$-axis nearly at the same
point, just below
$x=0.5$, while the $\overline {\rm MS}$ line crosses at above $x=0.5$. This is
understandable
since we saw from Eqs.(\ref{QSCI}, \ref{QNSMS}) that the first moment of
$\Delta
q_S^\gamma$ vanishes in the CI-like schemes while it is negative in the
$\overline {\rm MS}$ scheme. As $x\rightarrow 1$, we observe that the
$\overline{\rm MS}$, CI, and AB
lines  continue to increase while the OS line starts to drop. We also see that
the $\overline {\rm MS}$ and CI lines tend to merge and the AB line comes above
those
two lines. These behaviors are inferred from
Eqs.(\ref{NLOMSbar}-\ref{NLOAB}, \ref{NLOOS}).

Fig. 2  shows the $Q^2$-dependence of $\Delta q_S^{\gamma}(x,Q^2,P^2)$ in the
OS scheme
in units of $(3 N_f <e^2>\alpha /\pi){\rm ln}(Q^2/P^2)$. Three lines with
$Q^2=30,~ 50$ and
$100~ {\rm GeV}^2$ almost overlap in whole $x$ region except in the vicinity
of $x=1$.
Indeed we see from Fig.2 that in the OS scheme $\Delta q_S^{\gamma}$ beyond
the LO   behaves
approximately as the one obtained from the box (tree) diagram calculation,
\be
  \Delta q_S^{\gamma({\rm Box})}(x,Q^2,P^2)=(2x-1) 3 N_f
<e^2>\frac{\alpha}{\pi}~{\rm ln}~
\frac{Q^2}{P^2}~.
\ee

Concerning the non-singlet quark distribution $\Delta
q_{NS}^{\gamma}(x,Q^2,P^2)$,
we find that when we take  into account the charge factors, it falls on the
singlet
quark distribution  in almost all $x$ region; namely two ``normalized"
distributions
$\Delta {\widetilde q}_S^{\gamma}\equiv\Delta q_S^{\gamma}/<e^2>$ and
$\Delta{\widetilde q}_{NS}^{\gamma}\equiv \Delta
q_{NS}^{\gamma}/(<e^4>-<e^2>^2)$  mostly overlap  except at very small $x$
region. The situation is the same in all four factorization schemes.
This is attributable to the fact that once the charge factors
are taken into account, the evolution equations for both  $\Delta
{\widetilde q}_S^{\gamma}$
and $\Delta{\widetilde q}_{NS}^{\gamma}$ have the same inhomogeneous LO and
NLO
$\Delta K$ terms and the same initial conditions at $Q^2=P^2$.

In Fig. 3 we plot the gluon distribution $\Delta G^{\gamma}(x,Q^2,P^2)$
beyond the LO
in units of $(3 N_f <e^2>\alpha /\pi){\rm ln}(Q^2/P^2)$, with three
different $Q^2$
values. Note that it is factorization scheme-independent up to the NLO. We
do not see
much difference in three lines with different $Q^2$. This means the $\Delta
G^{\gamma}$
is approximately proportional to ${\rm ln}(Q^2/P^2)$, or $1/\alpha_s(Q^2)$~.
But, compared with quark distributions, $\Delta G^{\gamma}$
is very much small in absolute value  except at the small $x$ region.

\bigskip

In summary, we have studied the factorization scheme dependence of the
parton distributions
inside the virtual photon. The scheme dependence is clearly seen in the
$n=1$ moments
and the large $x$-behaviors of the quark distributions.
More details, together with the analysis on the scheme dependence of parton
distributions
near $x=0$, will be reported elsewhere.

\bigskip

%%%%%%%%%%%%%%%%%%%%%%%%%%%%%%%%%%%%%%%%%%%%%%%%%%%%%%%%%%%%
\vspace{0.5cm}
\leftline{\large\bf Acknowledgement}
\vspace{0.5cm}

We thank S.~J.~Brodsky, J.~Kodaira, M.~Stratmann 
and O.~V.~Teryaev for valuable
discussions. This work is partially supported by the Monbusho
Grant-in-Aid for Scientific Research NO.(C)(2)-09640345.
%%%%%%%%%%%%%%%%%%%%%%%%%%%%%%%%%%%%%%%%%%%%%%%%%%%%%%%%%%%%

\newpage

\newpage

%%%%%%%%%%%%%%%%%%%%%%%%%%%%%%%%%%%%%%%%%%%%%%%%%%%%%%%%%%%%
\vspace{0.5cm}
\leftline{\large\bf Figure Captions}
\vspace{0.5cm}
%%%%%%%%%%%%%%%%%%%%%%%%%%%%%%%%%%%%%%%%%%%%%%%%%%%%%%%%%%
\baselineskip 16pt
\noindent
\begin{enumerate}

\item[Fig. 1]\quad
Factorization scheme dependence of the polarized
singlet quark distribution
$\Delta q_S^{\gamma}(x,Q^2,P^2)$ to the NLO in units of
$(3N_f <e^2>\alpha/\pi){\rm ln}(Q^2/P^2)$ with $N_f=3$,
$Q^2=30~{\rm GeV}^2$, $P^2=1~ {\rm GeV}^2$, and
%the QCD scale parameter
$\Lambda=0.2~{\rm GeV}$,
%We plot the
for $\overline {\rm MS}$(dash-dotted line), CI (solid line),
AB (short-dashed line), and OS (long-dashed line) schemes.

\item[Fig. 2]\quad
The polarized singlet quark distribution
$\Delta q_S^{\gamma}(x,Q^2,P^2)$ to the NLO
in the OS scheme in units of $(3 N_f <e^2>\alpha
/\pi){\rm ln}(Q^2/P^2)$ with three different $Q^2$ values,
for $N_f=3$, $P^2=1~ {\rm GeV}^2$, and $\Lambda=0.2~ {\rm GeV}$.

\item[Fig. 3]\quad
The polarized gluon distribution $\Delta G^{\gamma}(x,Q^2,P^2)$ beyond the LO
in units of $(3 N_f <e^2>\alpha /\pi){\rm ln}(Q^2/P^2)$  with three
different $Q^2$ values, for $N_f=3$, $P^2=1~ {\rm GeV}^2$, and
$\Lambda=0.2~ {\rm GeV}$.

\end{enumerate}

%%%%%%%%%%%%%%%%%%%%%%%%%%%%%%%%%%%%%%%%%%%%%%%%%%%%%%%%%%%%
\newpage
\pagestyle{empty}
%\vspace*{-5cm}
\input epsf.sty
\begin{figure}
\centerline{
\epsfxsize=16cm
\epsfbox{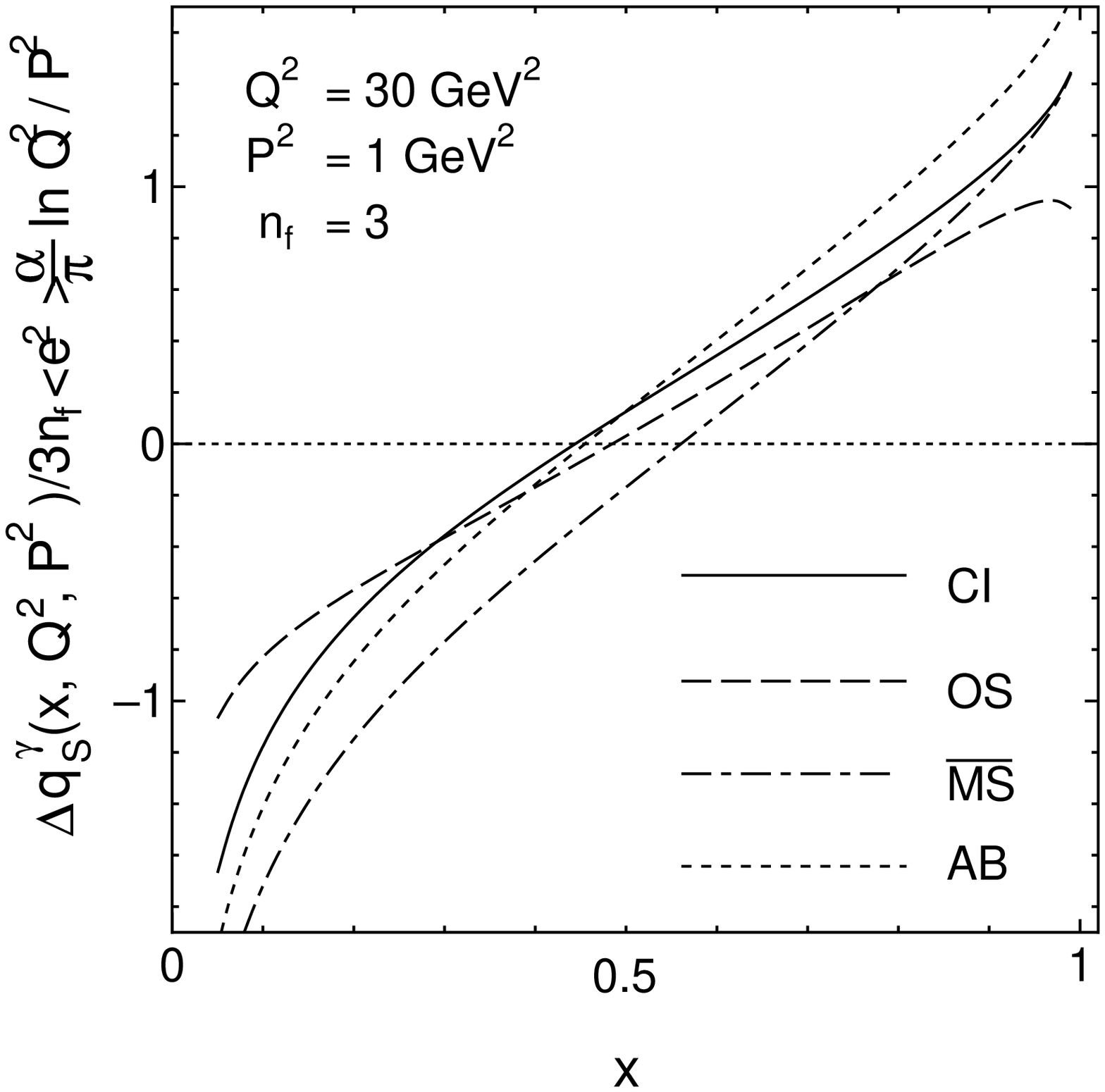}}
\vspace{-3.5cm}
\centerline{\large\bf Fig. 1}
\end{figure}

\newpage
\pagestyle{empty}
\input epsf.sty
\begin{figure}
\centerline{
\epsfxsize=16cm
\epsfbox{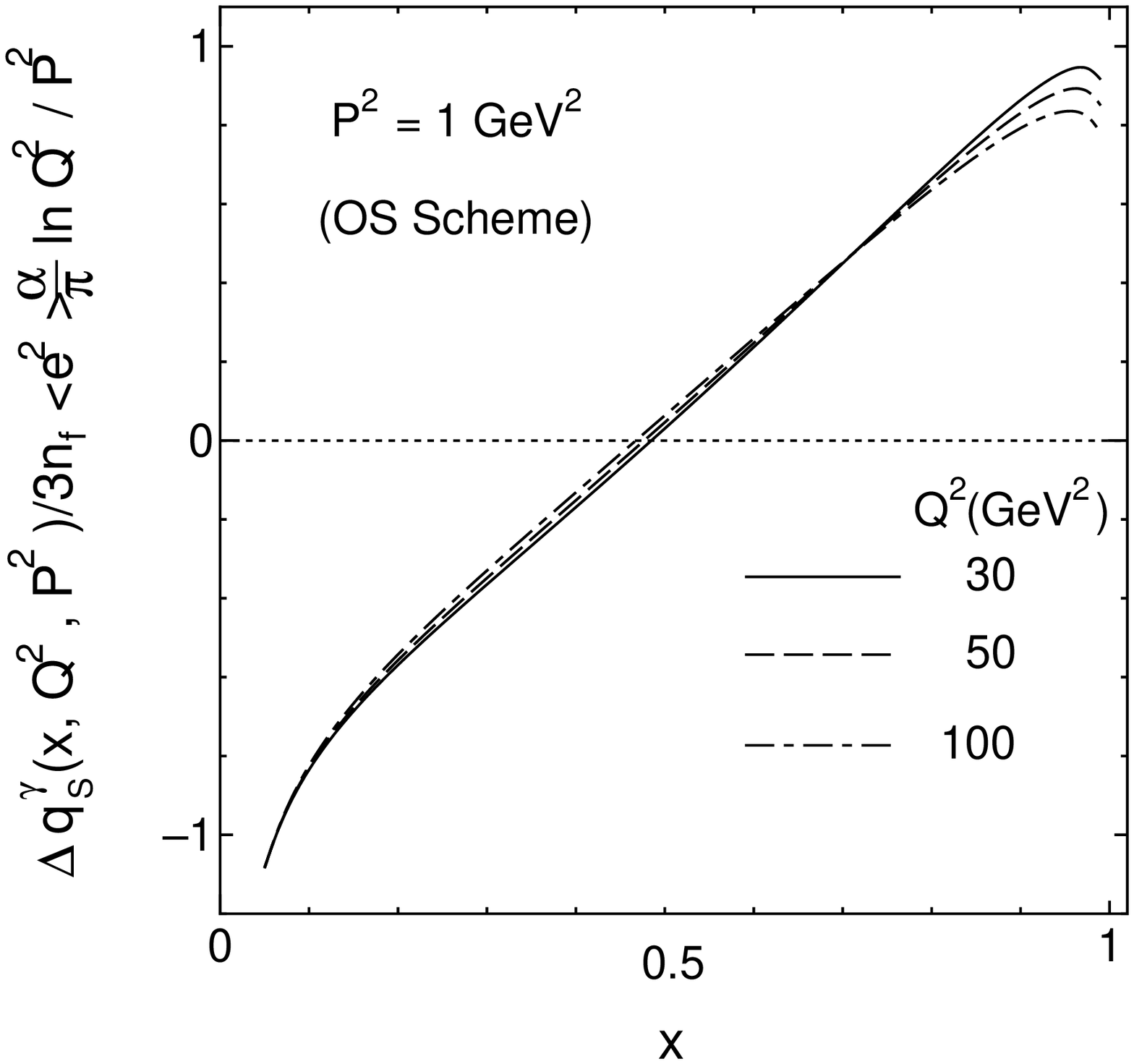}}
\vspace{-3.5cm}
\centerline{\large\bf Fig. 2}
\end{figure}

\newpage
\pagestyle{empty}
\input epsf.sty
\begin{figure}
\centerline{
\epsfxsize=16cm
\epsfbox{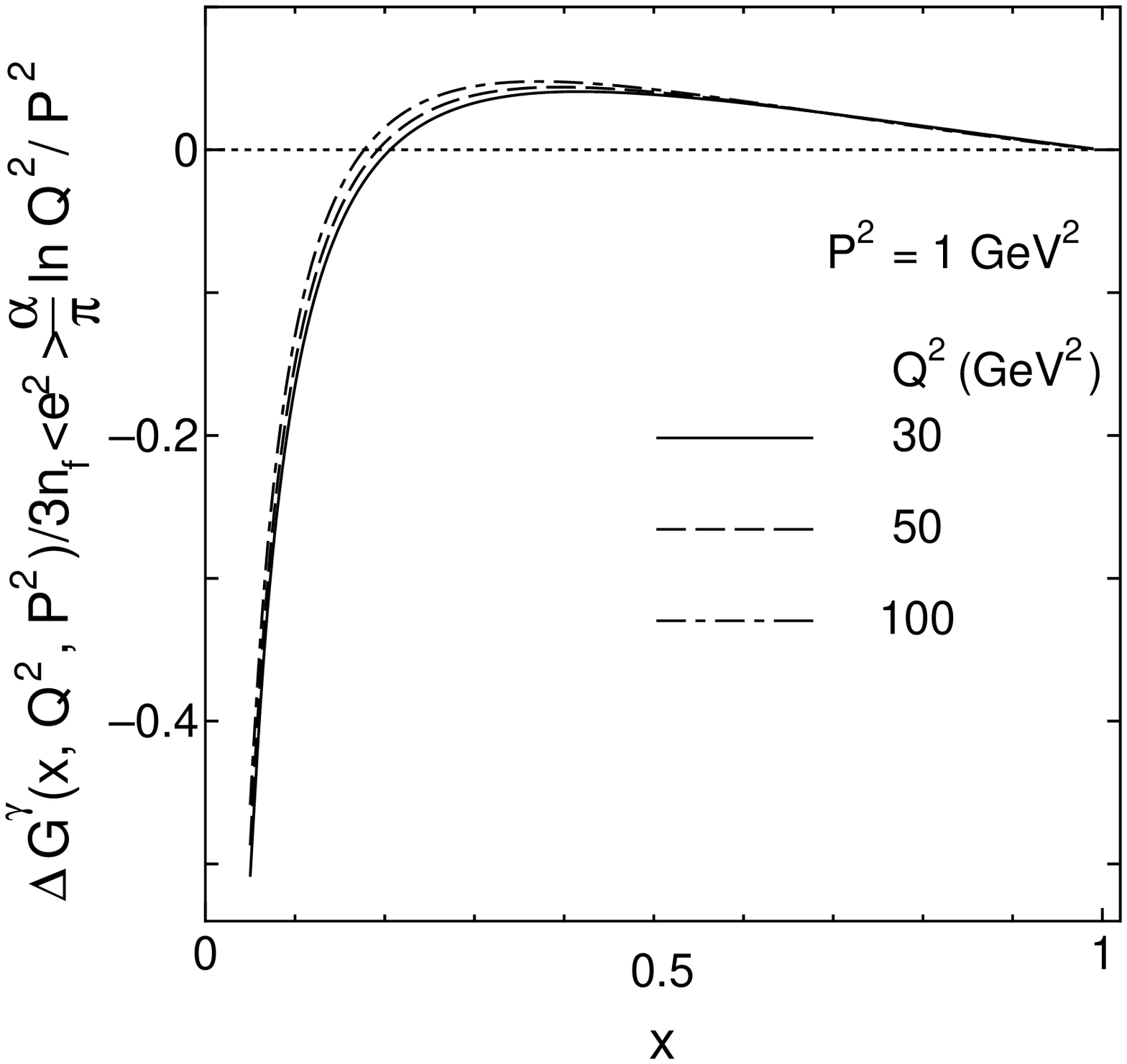}}
\vspace{-3.5cm}
\centerline{\large\bf Fig. 3}
\end{figure}

%%%%%%%%%%%%%%%%%%%%%%%%%%%%%%%%%%%%%%%%%%%%%%%%%%

\end{document}